\documentclass[pra]{revtex4}

 \usepackage{amsmath}
 \usepackage{graphicx}

\begin{document}

 \def\BE{\begin{equation}}
 \def\EE{\end{equation}}
 \def\BA{\begin{array}}
 \def\EA{\end{array}}
 \def\BI{\begin{itemize}}
 \def\EI{\end{itemize}}
 \def\BEA{\begin{eqnarray}}
 \def\EEA{\end{eqnarray}}
 \def\ra{\rangle}
 \def\la{\langle}
 \def\p{{\vec \rho}}
 \def\nn{{\nonumber}}
 \def\k{{\vec k}}
 \def\w{{\omega}}
 \def\W{{\Omega}}
 \def\r{{\vec r}}
 \def\Bl{{\Bigl(}}
 \def\Br{{\Bigr)}}
 \def\q{{\vec q}}
 \def\l{{\lambda}}
 \def\mb{{\mathbf}}
 \def\E{{\cal E}}


 \title{Multivariate quantum memory as controllable delayed multiport beamsplitter}

 \author{A.~ N.~ Vetlugin, and I.~V.~Sokolov}
 \address{St. Petersburg State University, SPbSU,\\
7/9~Universitetskaya~nab.,~St.~Petersburg,~199034~Russia}

 \begin{abstract}

The addressability of parallel spatially multimode quantum memory for light
allows one to control independent collective spin waves within the same cold atomic ensemble.
Generally speaking, there are transverse and longitudinal degrees
of freedom of the memory that one can address by a proper choice of the
pump (control) field spatial pattern. Here we concentrate on the mutual
evolution and transformation of quantum states of the longitudinal modes of collective spin
coherence in the cavity-based memory scheme. We assume that these modes are coherently
controlled by the pump waves of the on--demand transverse profile,
that is, by the superpositions of waves propagating in the directions,
close to orthogonal to the cavity axis. By the write-in,
this allows one to couple a time sequence of the incoming  quantized signals to
a given set of superpositions of orthogonal spin waves. By the readout, one can retrieve
quantum states of the collective spin waves that are controllable superpositions of the
initial ones and are on demand coupled to the output signal sequence. In general case, the memory
is able to operate as a controllable delayed multi-port beamsplitter, capable of transformation
of the delays, the durations and time shapes of signals in the sequence.

We elaborate the theory of such light--matter interface for the spatially multivariate
cavity-based off--resonant Raman--type quantum memory. Since in order to speed up the
manipulation of complex signals in multivariate memories it might be of interest to store
relatively short light pulses of a given time shape, we also address some issues of
the cavity-based memory operation beyond the bad cavity limit.

 \end{abstract}

 \maketitle

\section{Introduction}

The basic operations of quantum information protocols include creation,  transformation and
communication with non-classical states. In continuous variables quantum information one manipu\-la\-tes
quadrature components of light, collective spin coherence, etc. The appropriate tool for these
operations is provided by quantum memories \cite{Hammerer10, Bussieres13, NJP13}.

Atomic ensembles and solid state systems with multiple degrees of freedom are used
in order to extend the quantum memory capacity by applying the multimode
in time, frequency or spatial domain protocols. The inclusion of the quantized light signal
transverse degrees of freedom, as well as of non--collinear  configurations
of the interacting waves \cite{Vasilyev08, Tordrup08, Surmacz08, Vasilyev10},
allowed one to consider the quantum memory device as a quantum hologram.

In the spatially multimode memories the number of stored transverse modes
is limited by diffraction or even by the geometric aperture of atomic ensemble in
some configurations. There were demonstrated theoretically the off--resonant QND quantum
memory for images \cite{Vasilyev08}, and the spatially resolving off-resonant
Raman--type memory \cite{Surmacz08}. The quantum volume hologram \cite{Vasilyev10,
Vasilyev12} combines the Raman-type interaction with the non--collinear configuration
of interacting waves in analogy to classical volume hologram. The memories
based on the resonant interaction in $\Lambda$--schemes were extended to
spatially-multimode case for the fast and adiabatic \cite{Golubeva12, Tikhonov14}
modes of operation. Typically, by the forward-propagating retrieval the single--pass
schemes are less sensitive to diffraction and therefore are capable of achieving high
density of storage of transverse spatial modes. An extensive numerical analysis
\cite{Zeuthen11} of parallel $\Lambda$--type memory capacity was performed
in the paraxial approximation for a realistic shape of atomic ensemble.

The spatially resolving storage and retrieval of speckle--like patterns
was demonstrated \cite{Chrapkiewicz12} experimentally, the orbital angular momentum modes
were effectively stored and retrieved \cite{Nicolas14}.

An important resource allowing for the manipulation of multiple quantized signals  stored in quantum
memories is provided by  the use of classical control field with more complex spatial and temporal
structure \cite{Zhang13, Kalachev13,  Vetlugin14} as compared to  co-- and counter--propagating
 illumination with quasi--monochromatic wave.  The use of multi-atomic ensemble
which occupies a significant part of the cavity volume, as opposed to variety of single-atom
microcavity-based memories, makes it possible to achieve storage in multiple spatial modes of the
collective spin and provides sensitivity to the control field propagation direction.
By scanning the illumination angle \cite{Zhang13} or under the condition of continuous phase-matching
control across the control field beam \cite{Kalachev13} one can reproduce spatial and temporal
structure of a weak signal pulse. The 2D addressability of degenerate cavity--assisted off--resonant
memory by the illumination, close to orthogonal, allows for the retrieval on demand
of the signals stored both in transverse and longitudinal collective spin waves  \cite{Vetlugin14}.

In this paper we consider the operation mode of cavity--assisted off--resonant
memory allowing not only to store and retrieve  time sequences of  light signals,
but to perform an on--demand unitary ``in--out'' transformation of their quantized amplitudes.
This mode of operation might be of interest as a step towards essential transformation of
quantum states in the devices based on multivariate quantum memories.
In this regime the memory is able to operate as a controllable delayed multi-port beamsplitter,
capable of transformation of the delays, the durations and time shapes of signals in the sequence.
In order to achieve this, we assume that longitudinal modes of the collective spin coherence
of cold atomic ensemble are coherently controlled by the superpositions of pump waves
propagating in the directions, close to orthogonal to the cavity axis. This is equivalent
to  the use of control field with the on--demand modulated transverse profile.
First, by the write-in one can couple a time sequence of the incoming  quantized signals to a
given set of superpositions of orthogonal spin waves.  By the readout, one can retrieve
quantum states of the collective spin waves that are controllable superpositions of the
initial ones and are on demand coupled to the output signal sequence.

We elaborate the theory of such light--matter interface for the spatially multivariate
cavity-based off--resonant Raman--type quantum memory. Since in order to speed up the
manipulation of sequences of many signals in multivariate memories it might be of interest to store
relatively short light pulses of a given time shape, we also address the problem
of optimal choice of the control field time profile for the cavity-based memory operation
beyond the bad cavity limit.


 \section{Light--matter interaction with spatially structured control field}

\begin{figure}[h!]
 \begin{center}
 \includegraphics[width=80mm]{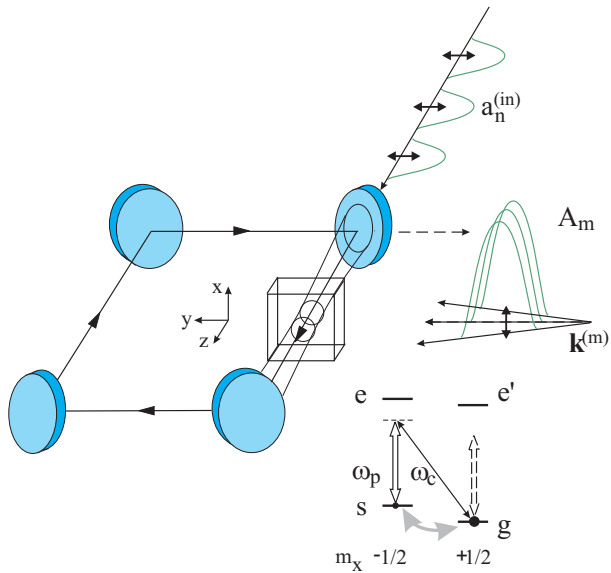}
 \caption{Quantum memory scheme.}
 \label{fig_1}
 \end{center}
 \end{figure}

We consider memory scheme scheme based on a high-Q single-port ring cavity of length L,
see Fig.~\ref{fig_1}.
The $y$-polarized cavity field is
 \BE
 \label{cavity_field_full}
  E(\mathbf r, t) =
  i\sqrt{\frac{2\pi\hbar\omega_c}{L}} a(t)\phi_0(x,y) \exp{\{i(k_cz-\omega_ct\}} + h.c.,
 \EE
where $a(t)$ is the quantized cavity mode slow  amplitude, that satisfies the standard
commutation relation, $[a(t),a^\dag(t)] = 1$, and the transverse distribution of the field
in \{x,y\} plane is given by the normalized zero-order Hermite-Gauss  profile $\phi_0(x,y)$,
 \BE
 \label{norm_transverse}
 \int dx\,dy |\phi_0(x,y)|^2 = 1.
 \EE
The empty cavity eigenfrequency is $\omega_c$, and  $k_c = \omega_c/c$.
The $y$-polarized input signal field, coupled to the cavity mode, at the input cavity mirror
($z=0$) is
 \BE
 E^{(in)}(x,y,0,t)=i\sqrt{\frac{2\pi\hbar\omega_c}{c}} a^{(in)}(t)\phi_0(x,y)
 \exp{\{-i\omega_c t\}} + h.c.,
 \label{input_field_definition}
 \EE
with the commutation relation for the input field slow amplitude of the form
$[a^{(in)}(t),a^{(in)\dag}(t')] = \delta(t-t')$.

The quantized field evolution in presence of the decay at $C/2$ rate and of the field
$a^{(in)}(t)$ at the input port is described by
 \BE
 \dot{a}(t) = -\frac{C}{2} a(t) +\sqrt C a^{(in)}(t), \quad
  a^{(out)}(t)=\sqrt C a(t)-a^{(in)}(t).
 \label{field_evolution_cavity}
 \EE
A spatially distributed ensemble of motionless atoms with angular momentum 1/2 both in
the ground and in the excited states is located inside the cavity, see Fig.~\ref{fig_1}.
We assume for simplicity that transverse dimensions
of the interaction area are determined by the cavity field waste, and the medium
length is $L_z$. The ground state collective spin coherence amplitude is given by
 \BE
  b(\mathbf r,t) = \frac{J_y(\mathbf r,t) +
  iJ_z(\mathbf r,t)}{\sqrt{2n_a\langle J_x^a \rangle}},
 \label{spin_amplitude}
 \EE
where $J_i(\mathbf r,t)$ for $i=x,y,z$ are the spin projection operators, and
the summation over atoms with the concentration $n_a$ is performed,
 $$
 J_i(\mathbf r,t)=\sum_a J_i^a(t)\delta(\mathbf r-\mathbf r_a).
 $$
The quantum memory regime implies that the number of atoms in the initial state
$J_x^a=+1/2$ remains almost unchanged. This allows one to substitute the operator $J^a_x$
with its mean value $\langle J_x^a \rangle$, and to arrive at the averaged over random
positions of atoms bosonic commutation relation of the form
 \BE
  \overline{[b(\mathbf r,t),b^\dag(\mathbf r\,',t)]}^a=\delta(\mathbf r -\mathbf r\,').
 \label{spin_commutator_desired}
 \EE
Due to the presence of constant magnetic field, oriented along $x$--direction, the
collective spin rotates at frequency $\Omega=\omega_{sg}$,
 \BE
  \dot{b}(\mathbf r, t) = -i\Omega b(\mathbf r,t).
 \label{spin_evolution_magnetic}
 \EE
Extending the addressable memory configuration of \cite{Vetlugin14}, we assume the pump
(control) field to be a controlled superposition of strong classical
$x$-polarized plane waves at frequency $\omega_p$, propagating in directions in the
\{$y,z$\} plane, close to orthogonal to the system axis $z$. Under the condition of the Raman
resonance, $\omega_c \approx \omega_p + \Omega$, and for large enough
magnetic splitting,  $\Omega\gg C/2$,  the cavity maintains quantized field
in the memory channel, see Fig. \ref{fig_1}, and suppresses the
entanglement channel. The last one is off-resonant to the cavity eigenfrequency.

Assuming a large frequency detuning from the electronic transition,
$|\omega_c-\omega_{eg}|\gg \Omega$, $C$, one can apply the effective QND Hamiltonian
\cite{Hammerer10}, that describes the light-matter interaction:
 \BE
 \label{eff_hamilt}
  H_{ef\!f} = \frac{2\pi\omega_c |d|^2}{(\omega_c-\omega_{eg})L}
  \int d{\bf r} J_z(\mathbf r,t) S_z(\mathbf r,t),
 \EE
where $d$ is dipole matrix element, $\omega_{eg}$ is the transition frequency,
$S_z(\mathbf r,t) = [A^\dag_r(\mathbf r,t)A_r(\mathbf r,t) -
A^\dag_l(\mathbf r,t)A_l(\mathbf r,t)]$ is $z$-projection of the Stokes vector.
Here $A_{r,l}$ are slow amplitudes of the right and left polarized fields
rotating in the $(x,y)$ plane, such that $E_{r,l}(\mathbf r,t) =
i(2\pi\hbar\omega_c/L)^{1/2}A_{r,l}(\mathbf r,t) + h.c.$ The pump field is composed
of N plane waves with the amplitudes $A_m(t)$, $m=1\ldots N$, propagating
in different directions in the $(y,z)$ plane. Hence, the atoms are interacting
with the field
 $$
 A(\mathbf r,t) = \sum_{m=1}^N A_m(t)
 \exp{\{i(k^{(m)}_z z+k^{(m)}_y y-\omega_pt)\}}\mathbf e_x +
 a(t)\phi_0(x,y) \exp{\{i(k_cz-\omega_ct\}} \mathbf e_y.
 $$
Here $k^{(m)}_z$ and $k^{(m)}_y$ are the wave vector projections for $m$'th pump wave.
Following the procedure described in more detail in \cite{Vetlugin14, Vetlugin13},
we retain the bilinear in $A_m(t)$ and $a(t)$ contributions to $H_{ef\!f}$, and find
the terms that come to the evolution equations
(\ref{field_evolution_cavity}, \ref{spin_evolution_magnetic}) in the Heisenberg picture
due to light--matter interaction. The contributions related to the entanglement
generation channel ($\dot a \sim b^\dag$, $\dot b \sim a^\dag$) are oscillating at
frequency $\pm 2\Omega$ and are averaged out on the time scale of interest $1/C$.

Note that we do not account here for atomic relaxation.
There are also omitted contributions to the effective QND Hamiltonian that
give rise to the cavity frequency correction due to the matter refraction index,
and to the atomic levels light shifts induced by the control field. In what follows we
imply that $\omega_c$ is the corrected cavity frequency, and the common time dependent
light shifts of the ground state sublevels mutually compensate.

The cavity and the collective spin evolution equations are then found to be
 \BE
 \label{field_from_spin_evolution}
 \dot{a}(t) = -\frac{C}{2} a(t) -
 \EE
 $$
 i Q\int d{\bf r}\,b({\bf r},t)\sum_{m=1}^N A_m(t)\phi_0^*(x,y) \exp\left\{i[(k^{(m)}_z-k_c)z+k^{(m)}_y y
 + \Omega t]\right\}  + \sqrt C a^{(in)}(t),
 $$
 \BE
 \dot{b}({\bf r}, t) = -i\Omega b({\bf r}, t) -  iQ\sum_{m=1}^N a(t) A^{*}_m(t)\phi_0(x,y)
 \exp{\{-i[(k^{(m)}_z - k_c)z + k^{(m)}_y y + \Omega t]\}},
 \label{spin_from_field_evolution}
 \EE
where
 $$
 Q=\frac{\pi\omega_c|d|^2}{\hbar  (\omega_c-\omega_{eg})L}\sqrt{2n_a\langle J_x^a\rangle}.
 $$

The evolution equations above describe the cavity field interaction with a set of independent
collective spin waves, controlled by the pump components, given the space profiles
that appear in (\ref{field_from_spin_evolution}, \ref{spin_from_field_evolution}) can
be considered as mutually orthogonal. To be specific, we assume that atomic distribution
is uniform within the cavity field waste, and $L_z$ is the medium dimension along $z$--axis.
The factors $\exp\{i[(k^{(m)}_z-k_c)z\}$ in (\ref{field_from_spin_evolution}) provide the
orthogonality given the $z$--projections of the pump wave vectors are chosen in such a way
that $(k^{(n)}_z - k^{(m)}_z) = 2\pi(n-m)/L_z$. For the pump waves propagating in the directions
close to $y$, when $|k^{(m)}_y|\approx |k^{(m)}|=\omega_p/c\gg |k^{(m)}_z|$,
the factor $\exp(ik^{(m)}_y y)$ can be substituted by $\exp[i(\omega_p/c)y]$ within
the cavity field waste, and the transverse profile of the spin waves involved
into interaction is given by $\phi_0(x,y)\exp[i(\omega_p/c)y]$.

We arrive at the following decomposition of the collective spin amplitude,
 \BE
  b(\mathbf r,t) = \sum_{m=1}^N b_m(t)\psi_m(\mathbf r)e^{-i\Omega t} + \ldots,
 \label{spin_decomposition}
 \EE
where $\psi_m(\mathbf r)$ are orthogonal spatial profiles of the
spin waves interacting with the cavity field,
 $$
 \psi_m(\mathbf r)=\frac{1}{\sqrt{L_z}}\phi_0(x,y)\exp{\{-i[(k^{(m)}_z-k_c)z+(\omega_p/c)y\}},
 $$
and ``\ldots'' stands for contribution of all other degrees of freedom of the
collective spin. The commutation relation (\ref{spin_commutator_desired}) yields
$[b_n(t),b_m^\dag(t)]=\delta_{nm}$. Inserting the decomposition
(\ref{spin_decomposition}) into (\ref{field_from_spin_evolution},
\ref{spin_from_field_evolution}), we reduce the evolution equations to the form
 \BE
  \dot{a}(t)=-\frac C2 a(t) +\sqrt C a^{(in)}(t) -i\sum_{m=1}^N k_m(t)b_m(t),
 \label{field_many_spins_evolution}
 \EE
 \BE
  \dot{b}_m(t)=-ik_m^{*}(t)a(t).
 \label{each_spin_field_evolution}
 \EE
The coupling parameter $k_m(t)=Q\sqrt {L_z}A_m(t)$ gives the state exchange
frequency between the $m$'th spin wave and the cavity field.

Consider time sequence of $N$ non--overlapping quantized input signals of
duration $\Delta T^{(W)}$ with the amplitudes $a^{(in)}_n$, such that
$[a^{(in)}_n,a^{(in)\dag}_m]=\delta_{nm}$, where $\{t_n\}$ are the
initial time moments of the signals,
 \BE
 \label{in_modes}
 a^{(in)}(t) = \sum_{n=1}^N a^{(in)}_n f(t-t_n) + \ldots, \qquad
 \int_{\Delta T^{(W)}}dt |f(t)|^2 = 1,
 \EE
where $f(t-t_n)$ are normalized time profiles of the temporal modes to be written,
and ``$\ldots$'' stands for the contribution of other temporal modes that ensure
correct commutation relations.

For the write--in of a single input signal $\sim a^{(in)}f(t)$ onto a single matter
oscillator on the (0,$T^{(W)}$) time interval, the Heisenberg--Langevin equations above
yield the following ``in--out'' relation,
 \BE
 \label{write_in_in_out}
 b(T^{(W)}) = K^{(W)}a^{(in)} + \sqrt{1-|K^{(W)}|^2}v^{(in)},
 \EE
where the write--in efficiency is $\eta^{(W)}=|K^{(W)}|^2\leq 1$, the bosonic amplitude
$v^{(in)}$ accounts for the initial vacuum contributions of the cavity field, the spin wave,
and other input temporal modes that are present by the $T^{(W)}$ time moment.
The problem of the optimal coupling parameter $k(t)$ control in order to achieve high
efficiency for the signal temporal mode of interest was thoroughly discussed in literature,
in what follows we shall address some aspects of such control beyond the bad cavity limit.

 \section{Memory as multi--port beamsplitter}

Assume that for the write--in of the n'th signal from the sequence, the coupling parameters
are taken in the form $k_m(t)= U^{(W)}_{nm}k(t-t_n)$, where $U^{(W)}$ is an arbitrary
unitary matrix. Inserting this to (\ref{field_many_spins_evolution},
\ref{each_spin_field_evolution}), one finds that the evolution equations reduce
to the case of interaction with a single spin oscillator with the amplitude
$b'_n(t)=\sum_m U^{(W)}_{nm}b_m(t)$.
Since before writing the signal sequence, the collective spin is supposed to be
in a vacuum state, the same is true
for all amplitudes $b'_n$. We also assume that between the write--in cycles
the cavity field restores its initial vacuum state due to finite cavity lifetime.
That is, we can apply the transformation (\ref{write_in_in_out}) incrementally for
$n=1\ldots N$ in the following form:
 \BE
 \label{beta_prime_out}
 b'_n(t_n+T^{(W)}) = K^{(W)}a^{(in)}_n + \sqrt{1-|K^{(W)}|^2}v'^{(in)}_n,
 \EE
where the unitary transform ensures that $[b'_n(t),b'^\dag_m(t)]=\delta_{nm}$.
To summarize, by the write--in of a time sequence of quantized input signals, one can
record their quantum state onto controllable superpositions of the collective spin waves.
In the initial basis for matter variables  this yields,
 \BE
 \label{beta_out}
 b_n(t_n+T^{(W)}) \sim K^{(W)}\sum_m (U^{(W)\dag})_{nm} a^{(in)}_m.
 \EE
Here and in what follows vacuum terms are omitted for brevity.

For the readout stage we assume representation
of the output field similar to (\ref{in_modes}), with the time duration $\Delta T^{(R)}$
of a retrieved signal from the sequence, and its time profile $g(t-t_n)$.
For the retrieval of a single output signal $\sim a^{(out)}g(t)$ from a single matter
oscillator on the (0,$T^{(R)}$) time interval, the evolution equations above
are solved to obtain $a^{(out)} \sim K^{(R)}b(0)$.

It is assumed that the time shape of the coupling parameter $k(t)$ during the
readout ensures efficient retrieval of the temporal mode $g(t)$.
For the n'th signal from the output sequence, the coupling parameters
are chosen to be $k_m(t)= U^{(R)}_{nm}k(t-t_n)$, where $U^{(R)}$ is unitary matrix.
As before, the evolution equations reduce to the case of interaction with a single
spin oscillator with the amplitude $b''_n(t)=\sum_m U^{(R)}_{nm}b_m(t)$.
The collective spin is supposed to be in the state, prepared during the
write--in, and the state of incoming quantized field is vacuum.
We also assume that between the readout cycles the cavity field restores
its initial vacuum state due to finite cavity lifetime.
We arrive at
 \BE
 \label{a_out}
 a_n^{(out)} \sim K^{(R)} b''_n(t_n) =  K^{(R)}\sum_m U^{(R)}_{nm} b_m(t_n).
 \EE

The complete ``in--out'' transformation of the input sequence of quantized
signals to the output one is composed of (\ref{beta_out}) and (\ref{a_out}),
and its essential part (apart from the vacuum contributions) is found to be
 \BE
 \label{a_to_a_final}
 a_n^{(out)} \sim K^{(R)}K^{(W)}\sum_m U^{(RW)}_{nm}a_m^{(in)},
 \EE
where $U^{(RW)}=U^{(R)}U^{(W)\dag}$. For $U^{(R)}=U^{(W)}$, when $U^{(RW)}=I$,
this yields $a_n^{(out)}=a_n^{(in)}$, but even in this simplest case the memory is able to
transform the delays, the durations, and time shapes of the signals in the sequence,
matching needs of next stage of the signals processing.
Given each row of $U^{(RW)}$ contains a unit in an arbitrary position, the
memory will change the signals order between the input and the output.
In general case, the described multivariate mode of operation makes it possible
to retrieve on demand superpositions of the input amplitudes. For instance, if
the complete transformation matrix $U^{(RW)}$ corresponds to 1:1 beamsplitter
($N=2$), an input sequence of two signals, squeezed in orthogonal quadratures, is
retrieved in the form of two entangled pulses. For $N>2$ one can simulate a multi-slit
interference of arbitrarily delayed and transposed signals with losses limited
by the memory efficiency.

As for experimental aspects of the proposed scheme, the effective shape of
an ultracold atomic ensemble may differ from the simple model we have used in order
to introduce orthogonal collective spin waves. For Gaussian distribution of atoms
along the system axis, there are more complex profiles of the pump field, as compared to
the plane waves with discrete values of the wave vector z-component $k_{iz}$, that interact
with orthogonal matter waves. As a suitable solution, one could use plane pump waves
with larger increment of $k_{iz}$. This would result in smaller overlap of the
relevant spin waves, in analogy to partially overlapping Glauber
coherent states. More detailed discussion of this issue is beyond the scope of this paper.

In order to effectively excite a high-quality cavity with active medium inside,
one has to minimize reflection of the input signal during the write-in. This can be achieved
by the impedance matching,  that is, by matching of the signal absorption by the medium placed
inside the cavity, with the cavity losses. It was demonstrated theoretically, that this approach
 works perfectly in the bad cavity limit for the atomic spin \cite{Gorshkov07},
 the atomic frequency comb \cite{Afzelius10}, and the controlled reversible
 inhomogeneous broadening \cite{Moiseev10} memories.

 Recent experimental demonstration  \cite{Bimbard14}  of the cavity--assisted
 cold atom memory with quasi--Gaussian control field pulses  was interpreted in
 terms of  adiabatic elimination of more general form  \cite{Stanojevic11}.

\begin{figure}[h!]
 \begin{center}
 \includegraphics[width=70mm]{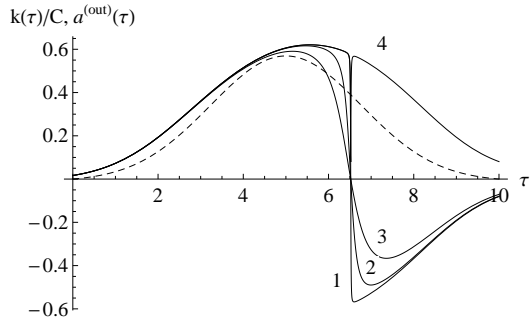}
 \caption{Plots of (dashed line) normalized Gaussian output signal $a^{(out)}(\tau)$ and (solid lines)
 several corresponding dimensionless coupling parameters $k(\tau)/C$ vs. time
 $\tau=Ct$ (that is,  time in units of the cavity lifetime). Here $T^{(R)}=10/C$, the readout efficiencies are $\eta^{(R)} =(1,4)$  0.958, (2)  0.957,  and (3) 0.946, respectively.}
 \label{fig_2}
 \end{center}
 \end{figure}

In the regime of multi--signal sequences, like in our proposal, it might be of interest
to speed up the manipulation of an elementary input signal beyond
the bad cavity regime in order to come to a suitable compromise between the speed and
the efficiency, which is limited also by other imperfections of the scheme. One can
achieve good efficiencies by matching the time shape of the light-matter coupling $k(t)$
with an on--demand smooth temporal mode of the signal $f(t)$ (by the write--in),
or $g(t)$ (by the readout) for the durations of the order of few cavity lifetimes
\cite{Vetlugin13, Kuzmin15}. It is worth noticing that the impedance matching approach
allows to find a set of the light--matter coupling parameter time profiles, that correspond
to almost equal values of the efficiency \cite{Kuzmin15}, as illustrated in fig.~\ref{fig_2}.
Having in mind the readout, in order to achieve a desired time shape of  the signal, it
might be needed to transfer some cavity excitations back to collective spin. Such transfer
occurs possible for a set of the control field and the collective spin amplitudes, given their
relative phase ensures absorption of the excess cavity photons.

%


 \section{Conclusion}

We have presented a detailed theory of the cavity assisted quantum memory with many longitudinal
modes of the collective spin coherence, controlled by the spatially structured classical
pump field. We have demonstrated that by the write--in in this regime of operation,
the memory is able to couple a time sequence of the incoming  quantized signals to a
given set of superpositions of orthogonal spin waves. By the readout, one can retrieve
quantum states of the collective spin waves that are controllable superpositions of the
initial ones and are on demand coupled to the output signal sequence. The memory
operation in the regime of controllable delayed multi-port beamsplitter, capable of transformation
of the delays, the durations and time shapes of signals in the sequence, might be of interest
for various tasks of quantum information. \vspace{2mm}

This work was supported by the Russian Foundation for Basic Research through the Projects 13-02-00254-a
and 15-02-03656-a. One of the authors (A. V.) acknowledges support by the Government of St. Petersburg
grant for young researches.

 \bibliographystyle{plain}
 
 \end{document}